# Local fractality: the case of forest fires in Portugal

M. Kanevski[a,*], Mário G. Pereira[b,c]

[a] Institute of Earth Surface Dynamics (IDYST), University of Lausanne, 1015 Lausanne, Switzerland. [Mikhail.Kanevski@unil.ch]

[b] Centre for Research and Technology of Agro-Environment and Biological Sciences, CITAB, University of Trás-os-Montes and Alto Douro, UTAD, Quinta de Prados, 5000-801 Vila Real, Portugal. [gpereira@utad.pt]

[c] Instituto Dom Luiz, IDL, Faculdade de Ciências da Universidade de Lisboa, Campo Grande, Edifício C8, Piso 3, 1749-016 Lisboa, Portugal [gpereira@utad.pt]

* Corresponding author: Mikhail Kanevski, Institute of Earth Surface Dynamics (IDYST), University of Lausanne, 1015 Lausanne, Switzerland. E-mail address: Mikhail.Kanevski@unil.ch Phone: +41 22 692 3531.

**Abstract**

The research deals with a study of local fractality in spatial distribution of forest fires in Portugal using the sandbox method. The general procedure is the following: (a) define a circle centred in each and all events with increasing radius $R$; (b) count the number of other events located within the circle of radius $R$, $N(R)$; (c) plot the growth curve which is the functional dependence of $N(R)$ versus $R$; and (d) estimate the local fractal dimension as the slope on $\log[N(R)]$ versus $\log[R]$. The computation is carried out by using the location of every fire event as a centre but without the final averaging over all the fires for a given $R$, which is usually performed to get a global fractal dimension and to estimate global clustering. Sandbox method is widely used in many applications in physics and other subjects. The local procedure has the ability to provide the most complete information regarding the spatial distribution of clustering and avoiding non-homogeneity and non-stationarity problems. Most of the analysis was performed using the National Mapping Burnt Area (NMBA) database which accounts for 32 156 fires during the 1975 – 2013 period. The results of local analysis are compared with a randomly generated pattern in forest zones (validity domain). The results demonstrate interesting local spatial patterns of clustering. Some results on global measures are reported as well.

# 1. Introduction

Clustering is an important technique in spatio-temporal point processes, monitoring networks analysis and preferential sampling, as well as in data mining topics in general (see, for example, Everitt et al. 2011; Illian et al. 2008; Kanevski and Maignan 2004; Tuia and Kanevski 2008). Depending on the problem considered, clustering studies have different origins and results' interpretation. For example, clustering of monitoring networks is closely related to dimensional (fractal) resolution of networks, in data mining to intrinsic dimensions and manifold learning and in spatial statistics to clustering of point processes. In the present research, clustering of events in space – forest fires in Portugal – are considered. A fractal based approach is applied in this study to quantify patterns of clustering. However, in order to avoid problems related to spatial stationarity and inhomogeneity, a local measure is used.

It is well known, that classical calculus and Euclidean geometry methods are not suitable to study fractals such as the irregular or fragmented shape of natural features, usually expressed by statistical scaling laws in the spatial – or time – domain which mainly results from the power-law behaviour of some real-world physical systems (Bunde and Havlin 1994; Falconer 2004; Lopes and Betrouni 2009; Sornette 2006). The main tool of fractal geometry is the fractal dimension (FD) or Hausdorff–Besicovitch dimension $D_h$ (Besicovitch 1964; Hausdorff 1919) which can be compared with the topological dimension $D_T$, both defined as following (Lopes and Betrouni 2009): The $D_T$ of a point is 0, of a curve is 1, of a plane is 2, and of an object in general Euclidean space $R^n$ is $n$; and, the Hausdorff–Besicovitch dimension is defined as $D_h=\ln(N)/\ln(1/r)$ where $N$ is the number of object's internal homotheties and $r$ the ratio of this homothety.

In most cases, $D_h$ cannot be easily computed. Lopes and Betrouni (2009) present, compare and classify the most used methods to estimate the FD into the following three groups: (*i*) Box-counting methods, including the Box-counting method, the differential box-counting method and the extended counting method; (*ii*) fractional Brownian motion methods, comprising the variogram method and the power spectrum and, finally, the (*iii*) area measurement methods, such as the Isarithm method, the blanket method and the triangular prism method.

Statistical analysis, including the assessment of temporal and spatial clustering properties, clearly contributes to a better general knowledge of the process, to rank and understands it's fundamental properties and drivers, as well as to assess the susceptibility, vulnerability, hazard and risk of natural and human disasters such as earthquakes and wildfires (Pereira et al. 2015; Telesca et al. 2007).

Global measures of clustering can be classified into three major categories (Golay et al. 2014; Tuia and Kanevski 2008): (1) topological indices, including distances between points, Voronoï polygons and Delaunay triangulation, where the frequency/area distribution of the polygons can be interpreted as an index of spatial clustering; (2) statistical indices, e.g. the Moran's index (Moran 1950), the Morisita index (Morisita 1959), the Ripley's K-function (Ripley 1977), the variance-to-mean ratio or the more recent multipoint Morisita index (Golay et al. 2014; Hurlbert 1990), and (3) fractal dimension which can be assessed by different methods such as the sandbox method (Feder 1988), the box-counting method (Russell et al. 1980; Smith et al. 1989), the lacunarity index (Mandelbrot 1982) and the Rényi's dimensions or information entropy (Orozco et al. 2013).

Most of the statistical indices and measures of fractal dimension are based on the counting of the number of events inside a changing size spatial domain which can be a circle of radius $R$ (Ripley's K-function, sandbox counting) a cell or a box (Morisita index, box-counting) followed by the estimation of the slope (fractal dimension) of the log-log dispersion plot of the number of events *versus* the size of the circle/cell. In the case of the lacunarity index, one possible approach is very similar but relies on a moving/gliding box, also of changing size, to assess the mass and the probability distribution and, therefore the lacunarity is defined as the ratio between the second and (the square of) the first moments of the probability distribution (Plotnick et al. 1996; Tuia and Kanevski 2008).

Examples of studies using different measures of fractal dimension to assess space, time or space-time clustering include, for example, the assessment of clustering properties in the seismicity activity in Switzerland using the m-Morisita index (Telesca et al. 2015), in the Southern Appennine Italian region with counting statistics (Lapenna et al. 1998), in California using detrended fluctuation analysis (Telesca et al. 2007) as well as for the clustering characterization of environmental monitoring networks, such as the Swiss Indoor Radon Monitoring Network (Golay et al. 2014). In all these examples, fractal or multifractal dimension was assessed for the entire spatial domain or dataset. However, it will be very interesting/important to assess the existence of potential clustering features at local scale. For this purpose, a measure of local fractality dimension (LFD) will be very welcome.

According to the European Forest Fire Information System (EFFIS), Portugal is the country most affected by fire incidence in the European context. In fact, according to the European Fire Database (EFD) provided by EFFIS, which includes the forest fire information (namely, the country's total burnt areas and number of fires) compiled by EU Member States and other European countries in the 1980 – 2013 period, Portugal is the country with the highest number of fires (hereafter, NF) and the second higher burnt area (hereafter, BA), respectively, with 29% of the total number of fires and 22% of the total BA in Europe. The statistics computed with the Burned Areas Perimeters (BAP) dataset for the 2000 – 2013 most recent period (**Figure 1**), also provided by EFFIS for Europe and North African countries of the Mediterranean basin, not only confirm this trend but even underscores the extent of this problem in Portugal in the more recent years, when Portugal presents the highest statistics of fire incidence, with 32% and 37% of total BA and NF, respectively (Pereira et al. 2014).

The intra-annual variability of the fire incidence is characterized by an annual cycle with a peak during summer centred in August and a secondary peak centred in March more evident for the north western regions (Amraoui et al. 2015; Pereira et al. 2013; Trigo et al. 2013). The intra-annual variability of the fire incidence is closely related to the Mediterranean type of climate of the country which is roughly characterized by a wet and mild semester promoting the growth of vegetation and a dry and hot semester leading to vegetation thermal and hydric stress as well as when most of the NF and highest BA are reported (Pereira et al. 2013; Pereira et al. 2017; Santos et al. 2015; Sousa et al. 2015). The inter-annual variability is dominated by the occurrence of two different kinds of atmospheric anomalous circumstances namely, (*i*) a *climate anomaly*, characterized by the existence of long dry periods in late spring and early summer and, (*ii*) a *weather anomaly*, the occurrence of very intense dry and hot spells in days of also extreme synoptic conditions (Pereira et al. 2005; Trigo et al. 2006).

The spatial distribution of the fire incidence discloses two different regions, in terms of fire density and heterogeneity, separated by the Tagus River (**Figure 1b**). The northern region is characterized by much higher fire density but also by more irregular topography, denser river network, higher concentration of coniferous forest and population density as well as a cooler and rainier weather (Pereira et al. 2011; Pereira et al. 2015). All these factors are important drivers of the fire incidence and the differences in the two regions, in terms of fire incidence statistics and fire danger factors, have suggested the existence of two different fire regimes in Portugal (Parente et al. 2016). Therefore, it is not surprising that the spatial patterns of the fire incidence (**Figure 1a**) also exposes the non-uniform spatial distribution of both NF and BA and the existence of clustering features easily assessed, even by visual examination.

The ability and robustness of the space-time permutation scan statistics (STPSS) were recently, tested for cluster analysis on aggregated datasets and then used to assess the existence (number, location, and space-time size) and statistical significance of detected clusters on the two Portuguese official databases (Pereira et al., 2015, Parente et al., 2016). The detected clusters were also characterized in terms of socioeconomic and environmental factors revealing that cluster areas are characterized by anomalous values of altitude, slope and vegetation cover while the fires belonging to the cluster occurred in days with also anomalous weather conditions affecting specifically the clusters' locations (Pereira et al., 2015, Parente et al., 2016). Another finding of these studies is that the obtained results of the cluster analysis were very dependent on several factors including fire density, dataset limitations as well as on the characteristics of the STPSS namely, the dimension of the scanning window. Therefore, the main objectives of this study are: (*i*) to propose a measure of local fractality in a close relationship with the concepts of validity domain and complete spatial randomness regarding its application; and, (*ii*) to assess local clustering features of forest fires in Portugal using a different new approach, namely the local fractal dimension, results which could be more independent on the selected method/techniques and dataset characteristics.

## 2. Description of case study

### 2.1. Study area

Portugal is the country located further to the southwest not only of Europe but also of the Iberian Peninsula (**Figure 1a**). The altitude spans from the sea level in the southern and western coast to about 2 000 m, in the north central region disclosing a relatively flat/homogenous territory at the south of Tagus River (**Figure 1b**) but much more heterogeneous landscape at the north where most of the mountain ranges may be found (Pereira et al. 2015).

The country has a temperate type of climate with dry and warm summer (*Csb*) in the north and dry and hot summer (*Csa*) in the south (AEMET 2011; Peel et al. 2007). This type of climate is driven by the latitude, the eastward general circulation, the Azores anticyclone, which defines the storm tracks reaching mainland and enhance the moderator effect of the Atlantic Ocean (at west), the continental effect of the Iberian Peninsula and the proximity to North Africa, more apparent during summer (Pereira et al. 2005; Trigo et al. 2006).

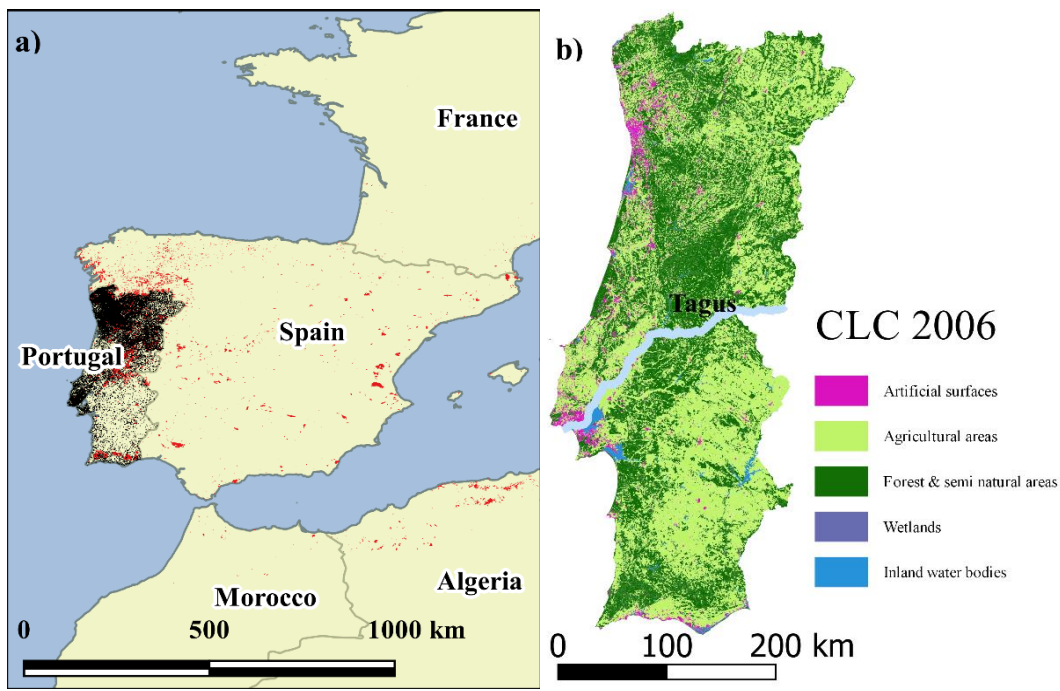

**Figure 1.** (a) The geographic location of mainland Portugal in the Southwest of the Iberian Peninsula, the spatial distribution of the National Mapping Burnt Area (NMBA) fire events in the 2001 – 2013 period (black dots), Burned Areas Perimeters (BAP) in Europe for 2000 – 2013

period (red polygons) and (b) the Corine land cover (CLC) for mainland Portugal in 2006 provided by the European Environment Agency (http://www.eea.europa.eu/).

According to the Corine Land Cover survey of 2006 (**Figure 1b**) provided by the European Environmental Agency (EEA 2013), the total area (89 000 km$^2$) of mainland Portugal is essentially and equally devoted to agriculture (47%) and occupied by forests and semi natural areas (48%) which, in turn, is dominated by scrublands (49%) and forests (47%) while open spaces with little or no vegetation only accounts for a minor fraction (4%).

### 2.2. The fire dataset

Portugal has two official fire datasets both provided by the Institute for the Conservation of Nature and Forests (ICNF). The Portuguese Rural Fire Database (PRFD) is a ground based dataset which has been updated over time (Parente et al. 2016; Pereira et al. 2011; Pereira et al. 2015). Each fire record in this dataset includes: (i) date and time of fire ignition and extinction; (ii) the BA by each fire in different land cover types (scrublands, forests and agricultural areas); (iii) a comprehensive set of additional information (e.g., fire cause, fire type), only available for specific sub periods; and, (iv) the spatial information, in the form of the name of the administrative regions (i.e., districts, counties and parishes) where the fire initiated for the 1980 – 2000 period and spatial coordinates of the fire events for the remaining 2001 – 2013 period.

On the contrary, the National Mapping Burnt Area (NMBA) is based on satellite imagery obtained in the end of the fire season and composed by the burnt area perimeters which comprises a detailed description of fire size and shape in an annual time scale. This dataset was recently reviewed to correct missing values and minor number of inconsistencies ending up with a final number of 46 000 fires for the 1975 – 2013 period (Parente et al. 2016). Accordingly, the PRFD is more appropriate when the analysis to be performed requires high/detailed temporal resolution, while NMBA should be selected when the knowledge of the location, shape and size of the burned area is of fundamental importance (Parente et al. 2016). Therefore, for this study, the NMBA was the selected fire dataset and the centroids of the BA polygons considered as the locations of the fire events (**Figure 2**). It should also be underlined that the NMBA have been used previously and recently in many studies such as to identify wildfire orientation burnt scars and its dependence with fire size (Barros et al. 2012; García-Portugués et al. 2014), to detect and characterize fire clusters (Parente et al. 2016), to map fire susceptibility, hazard and risk (Parente and Pereira 2016; Verde and Zêzere 2010), to assess land cover proneness to fire (Barros and Pereira 2014).

# 3. Methods

In this section a concept of validity domain, the reference patterns as well as global and local measures used in this study are shortly presented.

## 3.1. Validity domain

An important aspect to take into account when estimating global fractal dimension (GFD) and clustering assessment is the validity domain (VD) which is a well-known concept for instruments and techniques (Macko et al. 2001), equations (Scheid et al. 2005), models (Kahrs and Marquardt 2007) and theories (Saengkaew et al. 2006). In essence, the VD is determined by a set of geographical (e.g., topography, land use/land cover), political (e.g., administrative or country boundaries) or socioeconomic (economic activity, age, gender, habited or inhabited areas) constrains which define the complexity of the study area as well as by the characteristics of the monitoring network such as the number and location of observation points and finite size of measurements (Tuia and Kanevski 2008).

The existence of a VD, or their factors, decrease the dimensionality and the fractal dimension of the phenomenon meaning that the estimated GFD cannot be solely interpreted in absolute terms and/or in comparison with the standard values for homogeneous distributions (2 and 1 for GFD and Morisita index, respectively) but, and in essence, taking into account the VD of the phenomenon (Tuia and Kanevski 2008). This relative assessment can be performed by comparing the estimated clustering/fractal measures obtained for the observed samples (real points) and for a set of simulated samples of random points homogeneously (no clusters) generated/distributed within the VD.

This comparison should take into account the fluctuations caused by the randomness procedure and the finite number of observed points. In the latter case, the random samples are generated with the same number of points ($N$) as in the observed sample. In the former, several random samples (with size $N$) can be subsampled by bootstrapping from a large set of $M$ ($>>N$) points randomly generated in the VD. As a rule of thumb, M can be estimated, for a 2D spatial domain, as $M \approx (L_x \times L_y)/\delta^2$, where $L_x$ and $L_y$ are the spatial dimensions (in the X and Y directions) of the spatial domain and $\delta$ the spatial resolution of the phenomenon. Assessing the fractal dimension for those random samples also allows estimating the uncertainty and sensitivity to the VD and finite number of points. In the case of the forest fires in Portugal, the obvious VD is the forest and semi natural areas, where most (70%) of total fires occur. Therefore, random samples within this VD, called RND pattern, were produced with 32 156 points randomly distributed within the forest and semi natural areas. The RND pattern and validity domain as well as the observed forest fires, called FFP pattern, within the same VD are shown in Fig. 2.

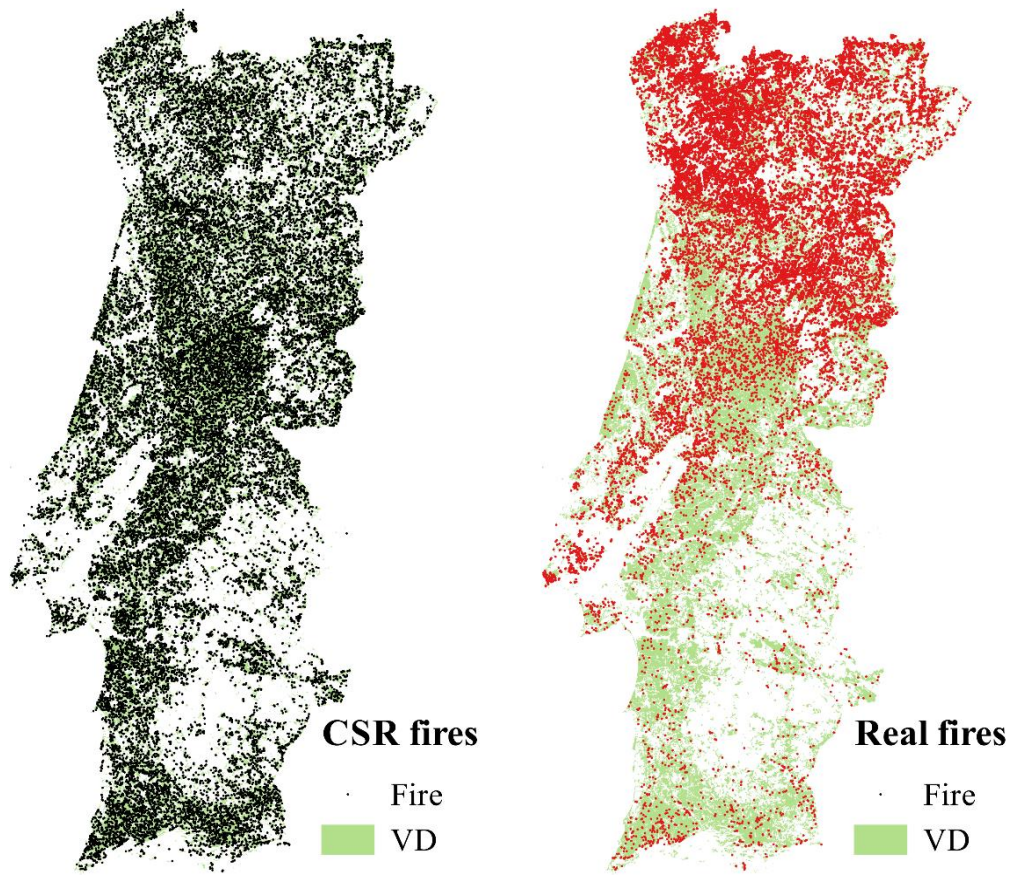

**Figure 2.** Locations of randomly generated points (left panel) and the fire polygon centroids (right panel) of the NMBA in the 1975 – 2013 period as well as the validity domain (VD) composed by forest and semi natural areas (green polygons) according to the Corine land cover 2006 inventory.

### 3.2. Preliminary exploratory spatial analysis & global fractal/clustering measures

Quantitative analysis can comprise different topological, statistical and fractal measures, both global and local (Kanevski and Maignan 2004; Tuia and Kanevski 2008). In the present research, the distribution between nearest neighbours is mainly considered. It is an important nonparametric tool used later in the selection of basic parameters when performing local fractal analysis. Global box-counting and sand-box counting measures are analysed to give an overview of the phenomena from different perspectives and to globally compare FFP and RND patterns.

The box-counting method was selected for being the most popular and frequently used to estimate the fractal dimension (Lopes and Betrouni 2009). Essentially, this method estimates the degree to which the pattern fills the space and is based on using a regular grid with cells/boxes of size $L$ superposed on the pattern under analysis and counting the number of cells $N(L)$, required to cover the entire pattern. Then the procedure is repeated using grids with different cells' size. For fractal patterns, the relationship between $N(L)$ and $L$ is a power law $N(L) \sim L^{-d}$, where $d$ is the fractal dimension. For the comparison purposes, it is important to note that fractal dimension is close to 2 if data occupy randomly two dimensional space, but deviates from 2 when random pattern is generated within the validity domain. This deviation characterizes the complexity of VD (Orozco 2015).

The sandbox method constitutes an efficient tool to estimate the fractal dimension (Tél et al. 1989). The method is based on circles of radius $R$ centred on each randomly selected event, $i$, or points belonging to the structure and counting the number of events (points or pixels) inside the circle, $R$. The fractal dimension $d$ is estimated from the growth curve $<N(R)> \sim (R)^d$ where $<N(R)>$ is a mean over all points with a fixed radius $R$ (Lopes and Betrouni 2009). Limitations and advantages of each of these methods are discussed in many previous studies (Dombrádi et al. 2007; Kun et al. 1995; Lopes and Betrouni 2009).

In order to study the points' distribution patterns in the space $(X,Y)$ but also of the phenomena itself – burnt area (BA) –, a functional measure of fractality/clustering can be applied (Lovejoy et al. 1987). In the present research, a functional sandbox counting method is used. In this case, the same sandbox counting measure is computed but this time only taking into account the points where the BA exceeds a given threshold $BA_{Thresh}$, i.e. sandbox growth curve is estimated only for the points where $BA(x,y) > BA_{Thresh}$. By changing $BA_{Thresh}$, the function is scanned and its patterns (spatial organization) can be revealed. The reference pattern for the functional measure is a shuffled data set, where all positions of the forest fires are the same, but the values of burnt area $Z$ are shuffled (randomized). This data set is called SHF. If there are no spatial structures in burnt areas (no correlations), changes in the clustering properties with the $BA_{Thresh}$ are only due to the reduced number of points. When there are spatial correlations, clustering changes significantly. The raw and SHF patterns will be compared using several thresholds and by visual inspection of their growth curves used to estimate dimension with the sandbox methods. This result is compared with the spatial correlogram computed after the (nonlinear) transformation of burnt area BA into Nscores – normally distributed scores. This transformation normalizes data and helps when data are highly skewed and have extremes and outliers, like in the case of forest fires.

### 3.3. Local sandbox-counting

The conventional sandbox method have been used to assess fractal or multifractal analysis in many different areas such as microvasculature in health and disease (Ward and Bai 2013), heavy metals contamination (Hakamata et al. 1998), spatial structure of population distribution (Orozco et al. 2013), drainage (Dombrádi et al. 2007) and natural river networks (De Bartolo et al. 2004), fracture network permeability (Jafari and Babadagli 2008), intermittency analysis of high-energy heavy-ion data (Kun et al. 1995) and in temperature-dependent properties of resistivity and thermoelectric power properties in bilayer films (Li et al. 2011; Tian et al. 1999).

In the conventional or global sandbox method $\langle N(R) \rangle \sim R^{df_{sand}}$, where $\langle N(R) \rangle$ means the average number of points inside the circle of radius $R$ centred on the events and $df_{sand}$ is the fractal dimension estimated with this technique. In the local sandbox method, $N(x, y; R) \sim R^{d(x,y)}$, which means no averaging and where the coordinates of the point process events $(x, y)$ are the locations where the circles of radius $R$ are centred. In this study, all events/points are considered as centres.

## 4. Results and discussion

### 4.1. Exploratory analysis & global fractal dimension

The preliminary exploratory analysis of the fire dataset reveals: (i) that the 32 156 fires in the 1975 – 2013 period burned a total area of 3 331 712 ha; (ii) a highly asymmetric distribution of the fire size (skewness=54), with BA ranging between 0.001 and 66 100 ha but with a median (average) of 22.1 ha (103.6 ha); and , (iii) an interquartile range , IQR=57.6 ha (standard deviation= 667 ha and a coefficient of variation equal to 6.4).

The visualisation of spatially distributed distances between first nearest neighbours for FFP and RND patterns are given in **Figure 3**. In **Figure 4**, a comparison between two distributions (cumulative probability density functions) are presented. This analysis gives important local and global information about point patterns. The study is nonparametric and does not depend on statistical hypotheses. It is evident that patterns are very different from this perspective.

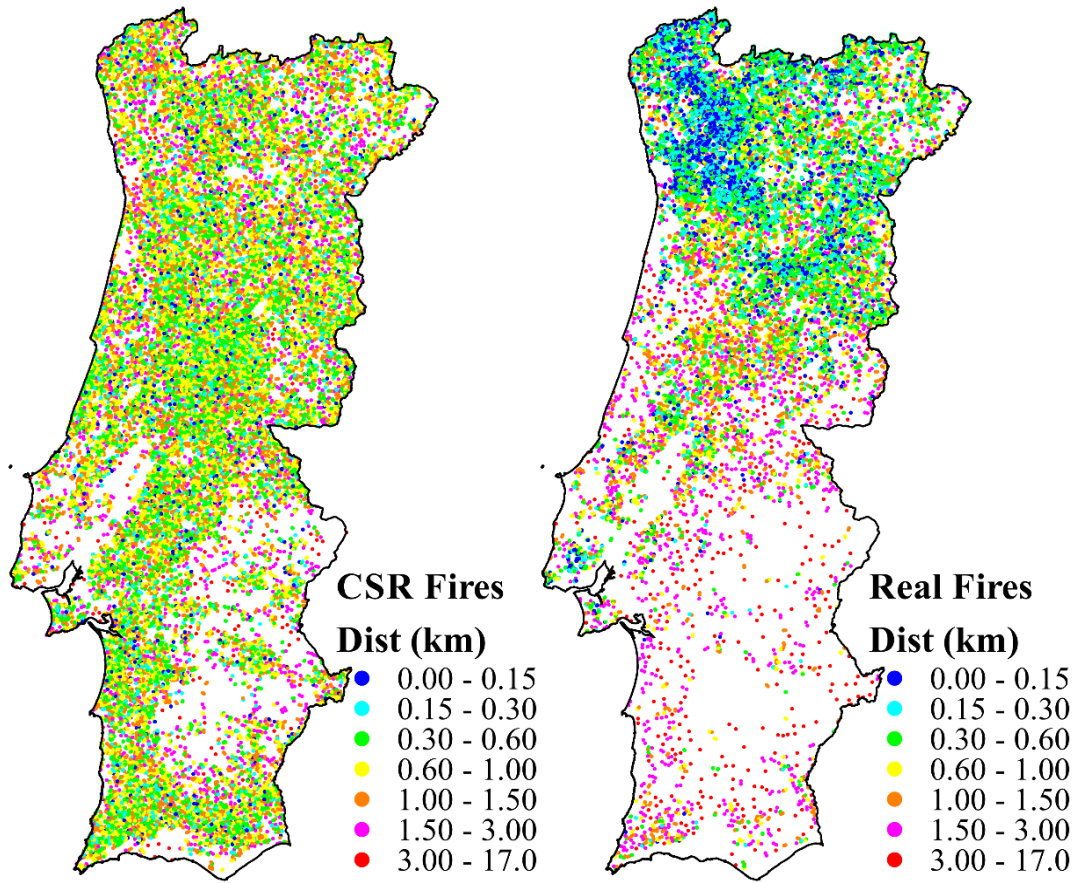

**Figure 3.** Spatial distribution of the distance between first nearest neighbours for complete spatial randomness (CSR) and real fires. Left – RND pattern, right – FFP pattern.

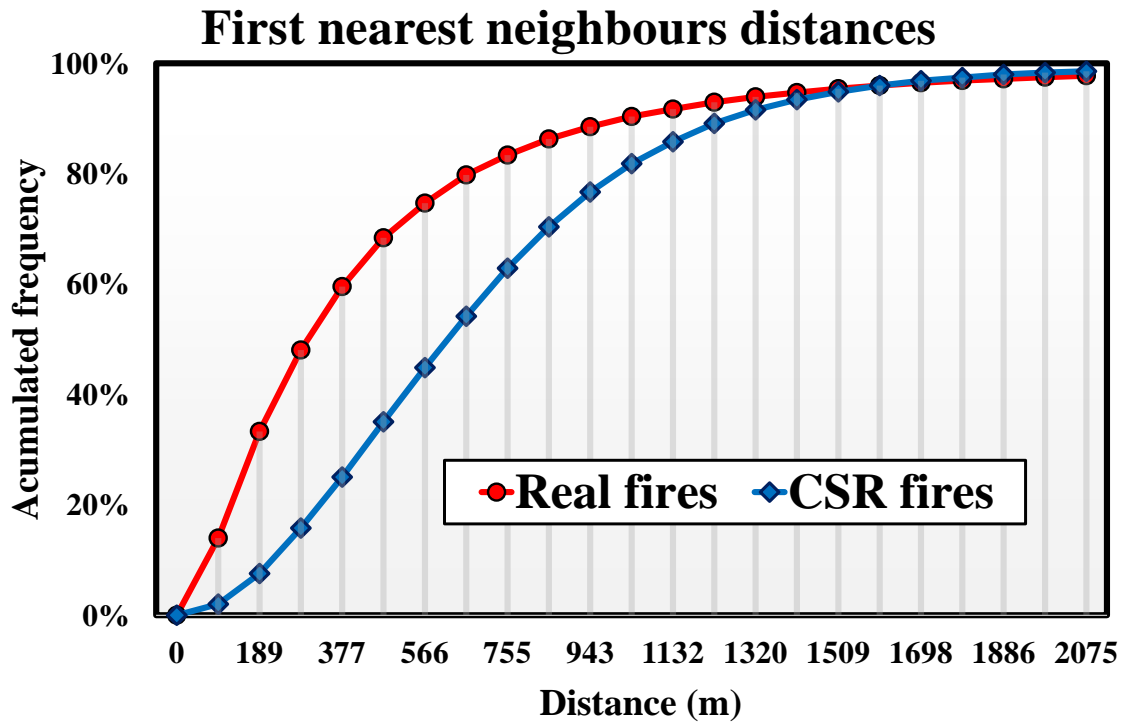

**Figure 4.** Comparison of cumulative probability density functions of the first nearest neighbours distances for real and complete spatial randomness (CSR) fires.

The estimation of the GFD with the box-counting and the sandbox methods provided slightly different results (**Figure 5**). Using the box-counting method, computed from 500 to 20 000 meters, for both the real and the randomly generated fires in the forested VD, the GFD presents a departure from exactly 2 due to the complexity of the validity domain, and its deviation from filled 2D space.

The box-counting growth curves were computed from L=500 meters to L=20 000 meters (**Figure 5**, left panel). The scaling region where the growth curves present a clear linear behaviour ($R^2 > 99.9\%$) extends from about 3 000 m to 20 000 m, more evident for random than for real fires, and with a GFD of 1.80 for CSR and 1.67 for real fires.

The sandbox growth curves were computed also from R=500 meters to R=20 000 meters (**Figure 5**, right panel). In this scaling region, the use of the global sandbox lead to the interesting observation that at the scale of about 3 000 m there is a break point and dimension/clustering for the forest fires is changing (from 1.56 to 1.82) in comparison with the RND pattern which gives almost the same dimension equalled to 1.88. It is interesting to mention that the same scale was observed with the functional measures of clustering and covariance function. Even if the computations have different meaning they evaluate the same phenomena from different perspectives.

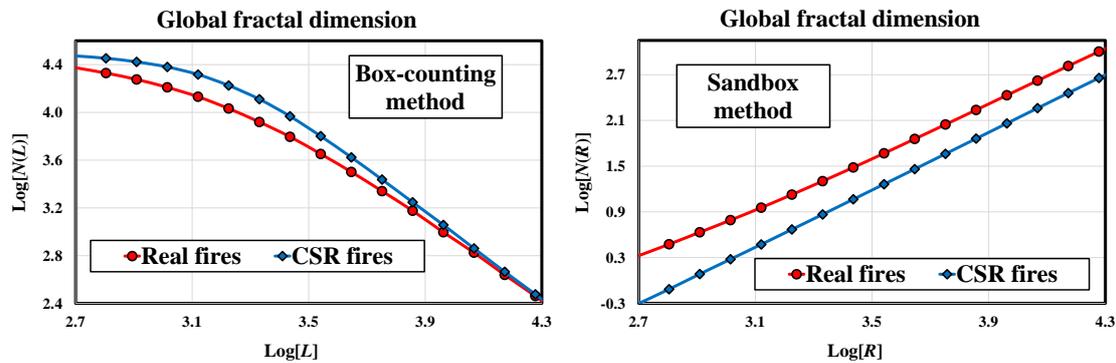

**Figure 5.** Box counting (left panel) and sand box counting (right panel) curves used for the global fractal dimension estimation.

The results of the functional sandbox-counting analysis of fires with BA above different thresholds were plotted with the results obtained with shuffled data set (**Figure 6**). The normalized growth curves obtained when considering all the fires and shuffled data are the same, however the differences between growth curves for real and SHF fires become apparent when considering fires with BA above increasing $BA_{Thresh}$, for smaller scales. This deviation is related to the structure in raw data compared with the shuffled data (no spatial structure). In general, if there are some structures in the data, clustering increases with the threshold. As it was already mentioned, for the shuffled data (without any structures) change of clustering is possible only due to the decreasing number of points. This effect can be taken into account by shuffling the data applying the thresholding procedure. In our case, there are enough data to see qualitative difference between the raw and shuffled data clustering with increasing the threshold. In **Figure 7**, the omnidirectional covariance function was computed for the raw and shuffled data. For SHF data set covariance equals to zero, while for raw data it has a clear structured part at small scales (around 3 000 m). Anisotropic covariance was also computed but only a minor dependence of the spatial correlations on the direction was observed.

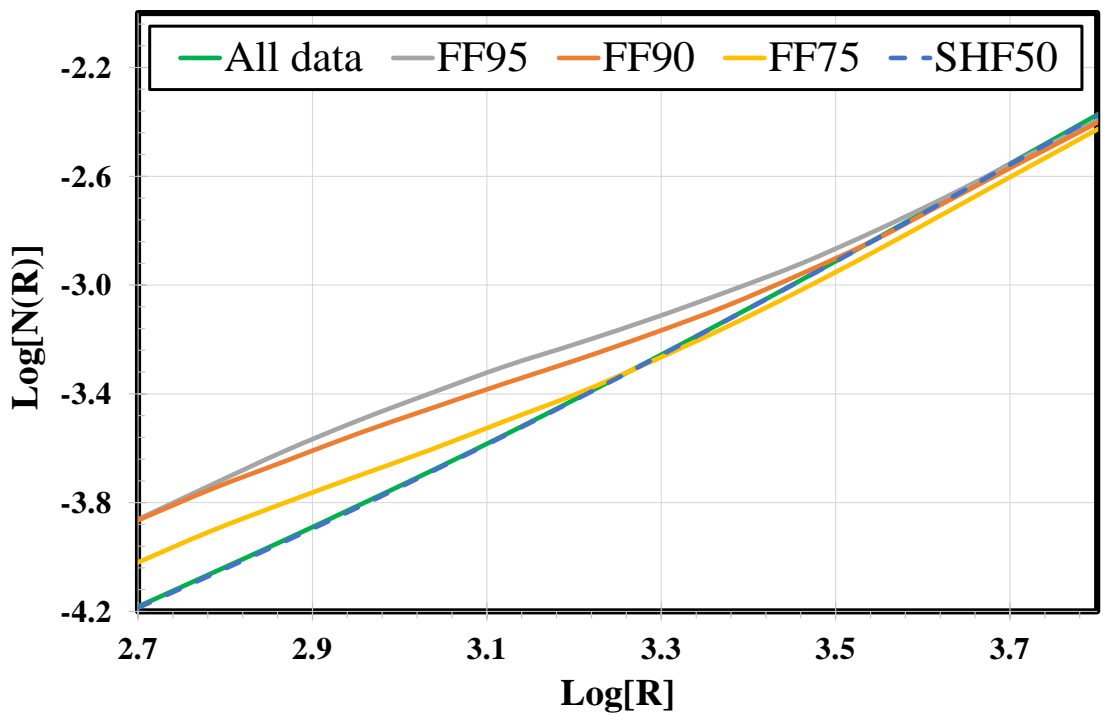

**Figure 6.** The sandbox-counting functional growth curves for forest fires and shuffled data.

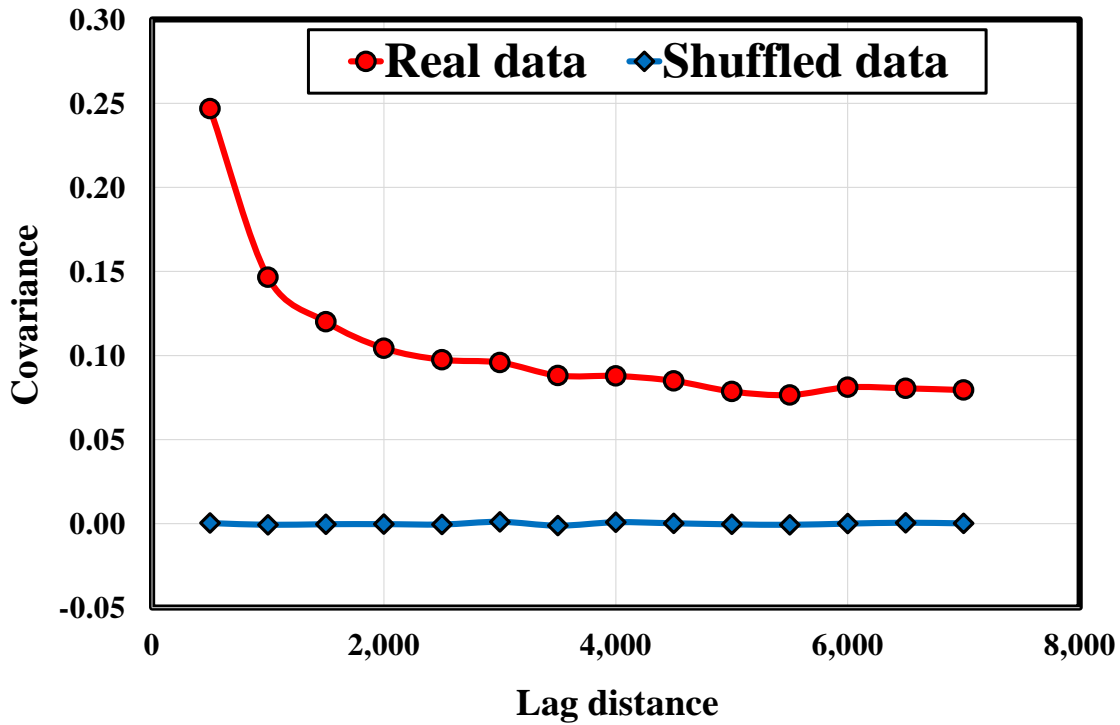

**Figure 7.** Spatial covariance functions computed for the real data (red filled circles) and shuffled data (blue diamonds). Covariance for the shuffled data equals to zero at all spatial lags (no spatial correlations). Functions were calculated on log-transformed data.

### 4.2. Local fractal analysis

Obtained results with the local sandbox method (**Figure 8**, top left panel) for the random fires reveal higher values of LFD (≥1.5) almost everywhere in the country except in the southern inland region, in the delta of the Tagus river and in metropolitan areas of the two largest cities of Portugal (Lisbon and Porto). This is a consequence of the fire generation and distribution process in the VD which leads to high LFD proportionally located where the density of forest and semi natural areas is also higher. On the other hand, for the real fires, high density of points with higher values of LFD is essentially restricted to the region at north of the Tagus river (**Figure 8**, top right panel). Since regions with high LFD are also more prone for clustering, this result is in very agreement with previous studies suggesting the existence of larger and most statistical significant space-time clusters of forest fires in the northern than in the southern parts of Portugal (Parente et al. 2016; Pereira et al. 2015).

The robustness of the LFD estimated with the sandbox for the real and CSR fires was also confirmed by the analysis of the variance of the regression fitting (**Figure 8**, bottom panels). In fact, low values of variance may be found at north of the Tagus river and, particularly, in the regions with higher average LFD. There is no apparent strong relationship between the variance and the local fractal dimension although smaller variability appears to be associated with also lower LFD (**Figure 8**).

Finally, to smooth the sandbox LFD map of real and CSR fires, the NF within circles were computed for increasing values of the radius ($NF_R$) which degrades the spatial pattern but helps to identify/filter the regions with higher clustering features and corresponds to the kernel density estimation (KDE). Results obtained for $R = 2.9$ km (**Figure 8**, central panels) reveal much higher $NF_R$ for real fires ($Q_2 = 28$ fires, Max = 148 fires) than for CSR fires ($Q_2 = 14$ fires, Max = 43 fires). In addition, regions with relatively higher $NF_R$ are wide spread in the VD for CSR fires while, for real fires, these regions may be found in the extreme northern part but essentially in the NW corner and, relatively smaller values are located over the north bank of the Tagus river. Obtained results are consistent with the findings of the clustering analysis performed previously for Portugal which detect a higher number of significant clusters precisely centred in these regions (Parente et al. 2016; Pereira et al. 2015). These results are also in very good agreement with the space-time Kernel Density Estimation (3D-KDE) of the detected spatial-temporal hotspots of forest fires in Portugal for the period 1990 – 2013, using the volume rendering technique (Tonini et al. 2016) .

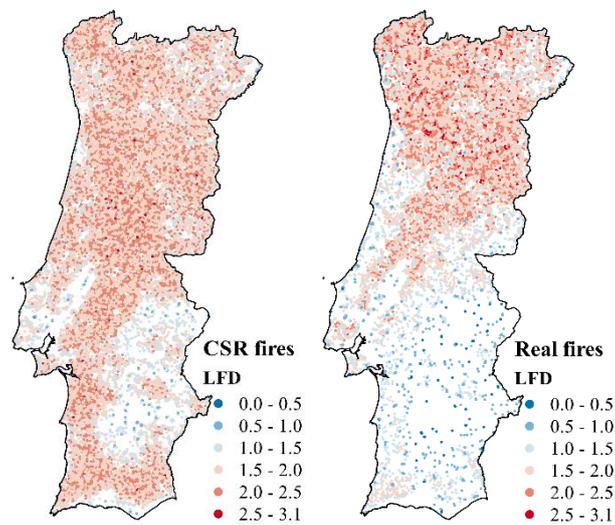

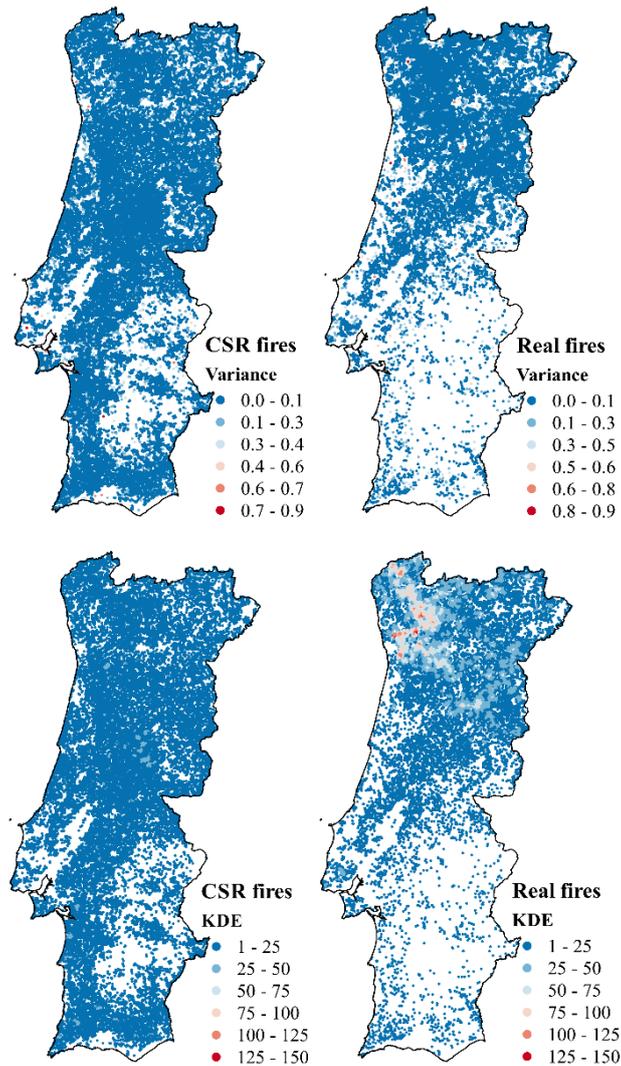

**Figure 8.** Local fractal dimension (LFD) computed with sandbox counting method for 1000 – 11 000 m and 10d samples (top panels), variance of the regression fitting (central panels) and the number of fires inside circles of radius $R = 2.9$ km (bottom panels) which is a kernel density estimation (KDE) measure, for the random (left panels) and real fires (right panels).

In **Figure 9**, the distributions of LFD for the forest fires and random patterns are presented. Their global characteristics are not very different in comparison with the spatial distribution of LFD (**Figure 8**, top right panel). Again, it means that global characteristics of fractality (clustering) can be very close despite of the very different local fractality behaviour.

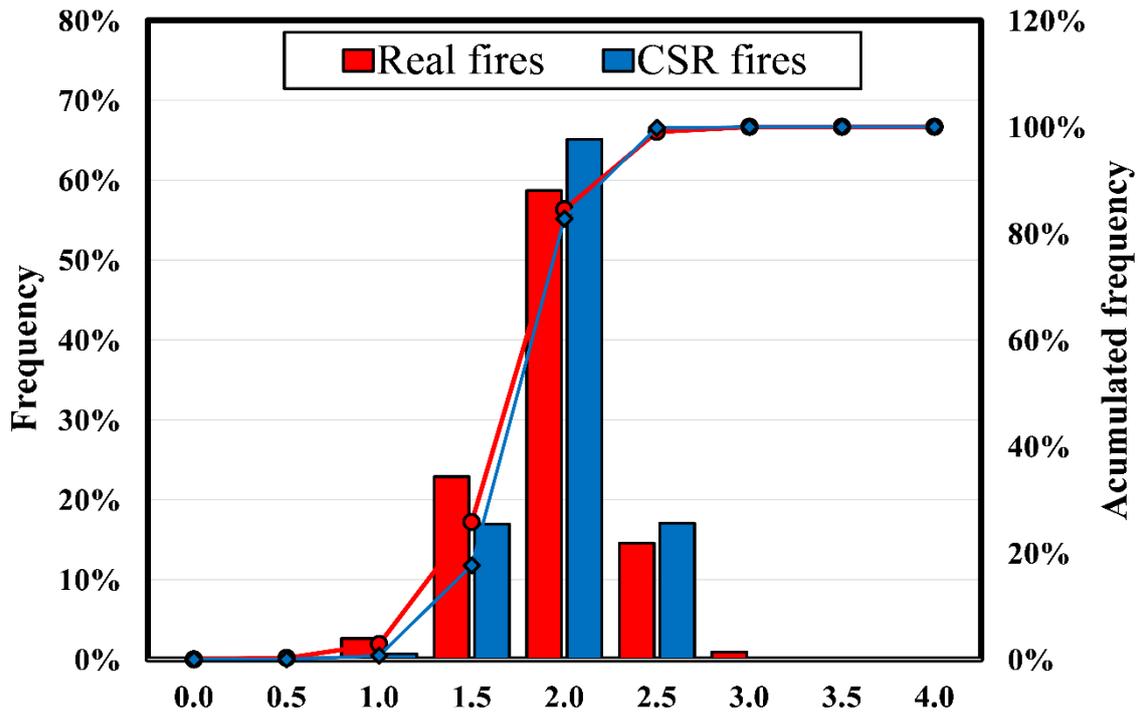

**Figure 9.** Distributions of local fractal dimensions LFD. Forest fires pattern (red): min LFD=0.00/0.14, meanLFD = 1.69, maxLFD = 3.15, skewnessLFD = -0.23, kurtosisLFD = 1.16. Random fires pattern (blue): minLFD = 0.00/0.39, meanLFD = 1.75, maxLFD = 2.69, skewnessLFD = -0.11, kurtosisLFD = 0.45.

## Conclusions

This study aimed to discuss the relationship between the concepts of fractal dimension, validity domain and complete spatial randomness as well as to purpose a new method to estimate local fractal dimension and to apply it to assess local clustering features of vegetation fires in Portugal. Major findings and conclusions can be summarized as follows.

The preliminary exploratory statistical analysis of the fire dataset reveal a heterogeneous spatial distribution of fires and a positively skewed fire size distribution which motivates to perform the spatial correlation analysis on log - transformed data and that discloses a slight anisotropy in the NW-SE direction associated with the predominant burnt scar orientation. Preliminary exploratory analysis also included studies of global fractal/clustering measures using two measures of global fractality (the box-counting and the sandbox counting methods), along with the concepts of validity domain and complete spatial randomness. The concept of validity domain which is defined as the region of interest within the study area, was used to properly interpret the relative fractal dimension or clustering indices, avoiding erroneous conclusions from the comparison between obtained absolute estimates for real data and standard measures for homogeneous distributions (fractal dimension = 2 for 2D CSR distributions). It is important to underline that this concept is rarely used except in more recent previous clustering analysis (such as Golay et al. 2014). The break point on the fractal dimension/clustering was detected at the scale of about 3 000 m with the sandbox method. The results seem to suggest the apparent greater difficulty of the conventional box-counting method to properly identify the particular clustering features, and the capacity of the global sandbox counting method to reveal change points in clustering.

The proposed method to estimate the LFD has been used/tested in a similar procedure and the local sandbox method was able to estimate the LFD for both samples of random and real fires. In the first case, the sandbox LFD pattern presents relatively higher values in almost the entire VD as higher density of fires is located in the regions with the largest area of forest and semi natural vegetation while in the case of real fires the pattern is substantially different with higher values concentrated almost exclusively located in the northern region, particularly in the NW corner of the country, and at the right margin of the river Tagus. The spatial pattern of the regression fitting variance confirms the robustness of the LFD estimates all over the VD while the number of fires within circles of increasing radius allow smoothing the LFD map and representing over densities of large vs medium vs small LFD.

It is also important to underline that the obtained results are in good agreement with the findings of previous cluster analysis respecting to the existence, number, location, size of statistically significant clusters identified for Portugal (Parente et al. 2016; Pereira et al. 2015; Tonini et al. 2016). The LFD and KDE spatial distributions are also consistent with the existence of two different fire regimes in Portugal, at the north and at the south of the Tagus river (**Figure 1b**) suggested by the fire incidence statistics and motivated by different natural (e.g., topography, precipitation patterns, vegetation cover) and anthropogenic (population density, socioeconomic activity) drivers (Parente et al. 2016).

Finally, we believe that the findings of this study will strongly contribute to a better understanding of the spatial patterns of fire regime in Portugal, particularly on some of the most important structural and circumstantial fire danger factors by which will allow better assess the risk of fire and practice a better management of fire, ecosystems and forest resources.

# Acknowledgements

This work was supported by: (i) the Herbette Foundation of the University of Lausanne; (ii) by the project Interact - Integrative Research in Environment, Agro-Chain and Technology, NORTE-01-0145-FEDER-000017, research line BEST, co-financed by FEDER/NORTE 2020; (iii) European Investment Funds by FEDER/COMPETE/POCI – Operacional Competitiveness and Internacionalization Programme, under Project POCI-01-0145-FEDER-006958 and National

Funds by FCT - Portuguese Foundation for Science and Technology, under the project UID/AGR/04033 and project FIREXTER PTDC/ATPGEO/0462/2014. We are especially grateful to ICNF for providing the fire data and to João Pereira for the final spelling and grammar review of the manuscript.